\documentclass[conference]{IEEEtran}

\usepackage{cite}
\usepackage{amsmath,amssymb,amsfonts}
\usepackage{graphicx}
\usepackage{textcomp}
\usepackage{booktabs}
\usepackage{url}
\usepackage{xcolor}

\usepackage{pgfplots}
\usepackage{pgfplotstable}
\pgfplotsset{compat=1.17}
\usepackage{algorithm}
\usepackage[noend]{algpseudocode}
\algrenewcommand\algorithmicrequire{\textbf{in:}}
\algrenewcommand\algorithmicensure{\textbf{out:}}
\makeatletter
\renewcommand{\ALG@beginalgorithmic}{\footnotesize}
\makeatother
\newcommand{\gemmqk}{\texttt{gemm\_qk}}
\newcommand{\gemmpv}{\texttt{gemm\_pv}}
\newcommand{\softmaxr}{\texttt{softmax}}
\newcommand{\vlmax}{\textup{\textsc{vlmax}}}
\newcommand{\Bq}{B_{q}}
\newcommand{\Bkv}{B_{kv}}

\newcommand{\chip}{EPAC chip}
\newcommand{\fpga}{FPGA}

\newcommand{\corr}{\boldsymbol{\rho}}

\definecolor{armStock}{HTML}{3FC8BC}
\definecolor{armScalar}{HTML}{2BA8C4}
\definecolor{armFused}{HTML}{3F6BC8}
\definecolor{armReor}{HTML}{5B2FA8}

\begin{document}

\title{Reorienting the Vector Axis: Restructuring FlashAttention for Long-Vector Processors}

\author{\IEEEauthorblockN{Emmanuel Irabor}
\IEEEauthorblockA{Barcelona Supercomputing Center\\
Barcelona, Spain\\
emmanuel.onyekachukwu@bsc.es}
\and
\IEEEauthorblockN{Marc Casas}
\IEEEauthorblockA{Barcelona Supercomputing Center\\
Universitat Polit\`{e}cnica de Catalunya (UPC)\\
Barcelona, Spain\\
marc.casas@bsc.es}}

\maketitle

\begin{abstract}
A Transformer is built around attention, and because every token attends to every other, the cost of that one operation grows with the square of the sequence length and comes to dominate the layer on long sequences. FlashAttention made that cost manageable by working through the attention matrix in tiles and updating the softmax as it goes. Almost all the work since has targeted GPUs.

Vector processors efficiently run High-Performance Computing (HPC) workloads. Modern commercial vector architectures feature large physical vector register files with hundreds of kilobytes of storage capacity. Because a single register holds up to 512 single-precision numbers, one instruction can execute 512 element operations, significantly reducing loop bookkeeping per result. These savings are proportional to how much of the register an instruction actually utilizes, making the choice of which loop to vectorize a first-order decision rather than a tuning detail. We port PyTorch's CPU FlashAttention kernel to a RISC-V CPU with 2KB vector registers and find that it achieves 5.17 of the 32 peak floating-point operations per cycle the core can issue (16.2\% of peak). The kernel walks each tile one query row at a time, leaving suboptimal dimensions to dictate vector instruction lengths. The product against $\mathbf{V}$ is vectorized along the head dimension—64 elements here, compared to the 512 elements a register holds. In the softmax, each query row requires a maximum and a sum: two reductions that collapse a full register to a single scalar once per tile. Additionally, because the hardware lacks a native exponential instruction, the softmax computes exponents via a polynomial whose coefficients compete with tile data for vector registers and cause spilling.

We propose two techniques. The first transposes the two matrices the kernel builds, the tile of scores and the running output, so that queries lie along the contiguous axis. This leaves the algorithm and its values unchanged but puts the vector length of every part of the kernel under one controllable parameter, the number of queries in a tile, so it can be chosen to match the machine's vector length. It also makes the maximum and the sum run along the query axis, which removes both reductions. The second moves the polynomial's coefficients into scalar registers, which stops the spilling and cuts the softmax region's vector memory instructions by 78.7\%.

On silicon the two together cut whole-kernel cycles by $1.84$ to $2.08\times$ at a head dimension of 64, raising floating-point utilisation from 16.2\% to 29.7\% of peak. An FPGA of the same design ranks the changes identically and scores them higher, at $2.71\times$, so the silicon figure is the conservative one.
\end{abstract}

\section{Introduction}
\label{sec:intro}

Vector processors utilizing Single Instruction Multiple Data (SIMD) paradigms offer superior throughput for High-Performance Computing (HPC) tasks. By executing a single instruction across multiple data streams simultaneously, SIMD reduces instruction-fetch overhead in the processor front-end while conveying extensive semantic context directly to the underlying hardware. These inherent efficiencies drove the dominance of early vector designs from the 1970s through the 1990s, highlighted by systems such as the CDC STAR-100~\cite{Schneck1987} and Cray-1~\cite{cray1}. Even as the 1990s saw the rise of massively parallel supercomputers containing thousands of nodes, vector-based execution retained its utility~\cite{limitations_vec,Krashinsky2004}, prompting chip manufacturers to integrate SIMD extensions into mainstream general-purpose microprocessors~\cite{sse:1999,3dnow:1999}. Today, as conventional performance scaling hits diminishing returns alongside skyrocketing demand for floating-point acceleration in modern workloads, vector processing is undergoing a major revival across both industry~\cite{Rico2017} and academia~\cite{Naithani2021,Sun2021,Domingos2021}.

This resurgence has propelled out-of-order (OoO) vector machines into mainstream commercial production, with NEC's SX-Aurora TSUBASA standing as a prime real-world implementation~\cite{aurora:2018,takahashi2023performance}. To maximize SIMD data-parallelism, vector registers now span a wide range, from the 512 bits of Arm's SVE in Fugaku~\cite{stephens2017sve,sato2020fugaku} to the $16\,384$ bits of NEC's SX-Aurora~\cite{komatsu2018sxaurora} and of RISC-V designs~\cite{minervini2023vitruvius,mantovani2026epac}.

Despite the relevance of vector architectures, FlashAttention~\cite{dao2022flashattention,dao2023flashattention2}, the standard implementation of attention, and almost all of the engineering that followed it, target GPUs~\cite{shah2024flashattention3,zadouri2026flashattention4}. This paper proposes several novel techniques to efficiently run FlashAttention on vector architectures featuring long register widths. What changes at this width is which architectural decisions matter. On a 128-bit unit, any loop with a handful of iterations fills the datapath, making the choice of which loop to vectorize a mere tuning detail. On a register holding 512 elements, however, a loop over a head dimension of 64 utilizes only an eighth of its capacity, and no instruction scheduler can recover the remaining throughput. PyTorch's CPU FlashAttention walks each score tile one query row at a time—a traversal that forces the key block to dictate the vector length of the score product and softmax, while leaving the head dimension to set the vector length for the value product. Recent RISC-V kernels retain one of these axes~\cite{titopoulos2026vectorized} or pack multiple heads into a single register once capacity exceeds the head dimension~\cite{gupta2026flashattentionv}; neither approach alters the underlying traversal.

This paper reorients the recurrence. We transpose the score tile and the running output, an outer-loop vectorisation~\cite{nuzman2008outer} of the query loop, so that the query axis becomes the contiguous axis of both. One tile parameter then sets the vector length of every region and can be set to the register width, and the per-query maximum and sum become vectors instead of reductions. The paper contributes:
\begin{itemize}
  \item a region-level characterisation, on silicon, of FlashAttention-2 on a $16\,384$-bit RISC-V processor that separates each region's vector-length limit from its access-pattern limit (Section~\ref{sec:char});
  \item a software exponential whose polynomial coefficients are held in scalar rather than vector registers, which removes spill traffic that a 32-register file otherwise makes unavoidable (Section~\ref{sec:softmax});
  \item reorientation of the recurrence, with an instruction-count model of when it wins (Section~\ref{sec:reorient});
        \item an evaluation on the EPAC chip and an FPGA that separates orientation
        from the softmax work: reorientation alone is worth $1.43\times$ in
        cycles (Section~\ref{sec:eval}).
\end{itemize}
None of this is a comparison against a GPU or a short-vector CPU, and none is needed for the claim, which is conditional: \emph{given} a long-vector machine and a query block longer than the head dimension, the orientation in which the recurrence is evaluated decides the vector length of every region, and with a query block as long as the register it removes about 30\% of the kernel's cycles at no change in the arithmetic performed.

\section{Background}
\label{sec:bg}

\subsection{Attention and FlashAttention tiling}
\label{sec:bg:fa}

Let $\mathbf{Q}, \mathbf{K}, \mathbf{V} \in \mathbb{R}^{N \times d}$ be the query, key and value sequences of a single attention head, where $N$ is the sequence length and $d$ the head dimension. The attention output $\mathbf{O} \in \mathbb{R}^{N \times d}$ is~\cite{vaswani2017attention}
\begin{align}
  \label{eq:attn}
  \mathbf{S} &= \alpha\,\mathbf{Q}\mathbf{K}^{\!\top}
                \in \mathbb{R}^{N \times N}, \nonumber \\
  \mathbf{P} &= \mathrm{softmax}(\mathbf{S})
                \in \mathbb{R}^{N \times N}, \nonumber \\
  \mathbf{O} &= \mathbf{P}\mathbf{V}
                \in \mathbb{R}^{N \times d}.
\end{align}
The softmax normalises each query against every key, so it runs along the second axis of $\mathbf{S}$, after subtracting each row's maximum to keep the argument in range. The scale $\alpha$ is a constant of the model, $1/\sqrt{d}$ here. It is applied as one multiply per score in every arm, so its numerical value affects no measurement below. Heads are independent, and FlashAttention runs them in parallel~\cite{dao2023flashattention2}, which makes a single head the unit of work. What follows measures one head on one core, the level at which vector behaviour is visible. Of the two matrix products one writes an $N \times N$ result and the other an $N \times d$ result, and Section~\ref{sec:char} shows that the difference is what decides their vector length.

Forming $\mathbf{S}$ explicitly costs $\Theta(N^2)$ memory. FlashAttention~\cite{dao2022flashattention} avoids it by tiling both operands, processing queries in blocks of $\Bq$ rows and keys in blocks of $\Bkv$, and carrying the normalisation of each query block as a running maximum and sum that are rescaled whenever a new tile moves the maximum~\cite{milakov2018online}. The tile shape $(\Bq, \Bkv)$ is a parameter of the implementation, not of the mathematics.

\subsection{Long-vector processors}
\label{sec:bg:longvec}

A vector instruction applies one operation to many elements. Short-vector SIMD units fix that number in the encoding, at four or eight elements, so a loop of a handful of iterations already fills the datapath. Vector processors instead carry the number in a register, set at run time, and recent designs have made it large: one register now holds several hundred values. The gain is that a single instruction stands for all of that arithmetic, so the instruction stream is short relative to the work it represents. The cost is that the gain is proportional to how much of the register an instruction actually uses.

The RISC-V vector extension is length-agnostic~\cite{riscvv2021}. Throughout, \vlmax{} is the largest number of elements a register holds, 512 in FP32 here, and $vl$ the active length of an issued instruction. With registers this long the scalar core must supply addresses, loop bookkeeping and configuration changes fast enough to keep a very wide back end busy. Decoupled vector architectures~\cite{espasa1996decoupled,asanovic1998vector} let it run ahead, but the supply problem returns whenever $vl$ is low, because each instruction then buys less work.

The processor studied here, EPAC~\cite{mantovani2026epac}, has registers of $16\,384$ bits. That is $32\times$ the register of current commercial RVV silicon~\cite{sifive2022x280} and of A64FX~\cite{sato2020fugaku}, at the long end of open RVV designs~\cite{cavalcante2020ara,perotti2022newara,cavalcante2022spatz}, and equal to the $256 \times 64$-bit registers of NEC's SX-Aurora~\cite{komatsu2018sxaurora}. Section~\ref{sec:method:target} describes it.

\section{Characterising the Stock Kernel}
\label{sec:char}

This section measures where the stock kernel's cycles go on the chip, region by region, and separates two limits that a whole-kernel number does not distinguish. The score product runs at the full register width and is held back by its access pattern. The value product has no access-pattern problem and is held back by its vector length. Each calls for a different remedy, and the two that follow are the subjects of Sections~\ref{sec:softmax} and~\ref{sec:reorient}.

Our starting point, the \emph{stock kernel}, is the FlashAttention-2~\cite{dao2023flashattention2} forward pass of PyTorch's CPU backend~\cite{paszke2019pytorch}: its blocking, loop order, data layout and online-softmax recurrence, unchanged. We start from FlashAttention-2 because the versions after it are specific to GPUs: FlashAttention-3~\cite{shah2024flashattention3} is built on warp specialisation and asynchronous copy engines, and FlashAttention-4~\cite{zadouri2026flashattention4} on tensor cores whose throughput has outgrown their exponential units. Neither mechanism exists on a vector unit. PyTorch has no vector backend for RVV, so we ported the kernel by implementing the vector primitives it calls with RISC-V vector intrinsics, keeping their structure: FP32 inner kernels, a polynomial approximation of the exponential because the hardware has none, and the row maximum and row sum as vector reductions. Algorithm~\ref{alg:natural} gives the traversal.

\begin{algorithm}[b]
\caption{Natural orientation (FlashAttention-2 forward).}
\label{alg:natural}
\small
\begin{algorithmic}[1]
\Require $\mathbf{S}_{ij}$ is $\Bq \times \Bkv$;\;
         $\mathbf{m}_i, \boldsymbol{\ell}_i$ hold one scalar per query row
\For{each query block $i$}
  \State $\tilde{\mathbf{O}}_i \gets 0$;\;
         $\mathbf{m}_i \gets -\infty$;\;
         $\boldsymbol{\ell}_i \gets 0$
  \For{each key block $j$}
    \State $\mathbf{S}_{ij} \gets \mathbf{Q}_i \mathbf{K}_j^{\!\top}$
          \Comment{$vl = \Bkv$}
    \State $\mathbf{m}' \gets \max\bigl(\mathbf{m}_i,
            \mathrm{rowmax}(\alpha\mathbf{S}_{ij})\bigr)$
          \Comment{$\Bq$ \emph{reductions}}
    \State $\corr_i \gets \exp(\mathbf{m}_i - \mathbf{m}')$;\;
           $\mathbf{m}_i \gets \mathbf{m}'$
          \Comment{$\Bq$ scalar $\exp$}
    \State $\mathbf{S}_{ij} \gets
            \exp(\alpha\mathbf{S}_{ij} - \mathbf{m}_i)$
          \Comment{$vl = \Bkv$}
    \State $\boldsymbol{\ell}_i \gets \corr_i \boldsymbol{\ell}_i +
            \mathrm{rowsum}(\mathbf{S}_{ij})$
          \Comment{$\Bq$ \emph{reductions}}
    \State $\tilde{\mathbf{O}}_i \gets
            \corr_i \tilde{\mathbf{O}}_i + \mathbf{S}_{ij}\mathbf{V}_j$
          \Comment{$vl = d$}
  \EndFor
  \State $\mathbf{O}_i \gets \tilde{\mathbf{O}}_i / \boldsymbol{\ell}_i$
        \Comment{$vl = d$}
\EndFor
\end{algorithmic}
\end{algorithm}

\subsection{Where the cycles go}
\label{sec:char:budget}

\begin{figure}[t]
  \centering
  \includegraphics[width=\columnwidth]{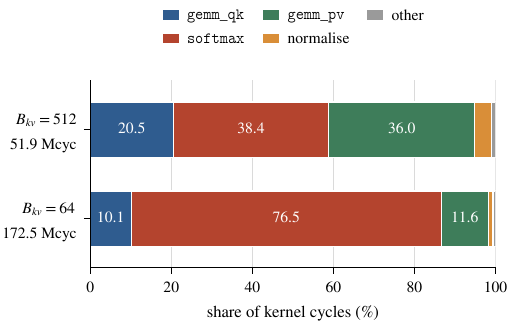}
  \caption{Cycle budget of the stock kernel by region, on the \chip{} at
           $N = 1024$, $d = 64$, $\Bq = 128$.}
  \label{fig:budget}
\end{figure}

Figure~\ref{fig:budget} gives the stock kernel's cycle budget at $N = 1024$, $d = 64$, in FP32 on one core of the \chip{}, with the methodology of Section~\ref{sec:method}. The first bar is the kernel at its own fastest tile, $\Bq = 128$, $\Bkv = 512$. No single region owns it: the softmax takes 38\%, the value product 36\% and the score product 21\%. Whole-kernel throughput is 5.17\,FLOP/cycle against a peak of 32.

The second bar holds $\Bq = 128$ and shortens the key block to $\Bkv = 64$. Whole-kernel cycles rise by a factor of 3.32, and \softmaxr{} accounts for 93\% of the increase. A region that grows more than sixfold when a block size shrinks, beside one that moves by a few percent, is not doing more work. It is issuing the same work at a shorter vector length, and the two subsections below separate that from the access-pattern effect.

The two matrix-product regions are limited by different things and call for distinct remedies: \gemmqk{} forms $\mathbf{Q}_i\mathbf{K}_j^{\!\top}$ and reads one operand in transposed orientation~\cite{goto2008anatomy}, while \gemmpv{} forms $\tilde{\mathbf{P}}_{ij}\mathbf{V}_j$, an untransposed product that vectorises along $d$.

\subsection{Wide vectors, strided operands: the score product}
\label{sec:char:stride}

\gemmqk{} sweeps the key axis, so $\Bkv$ and not $d$ sets its vector length. At $\Bq = 128$, $\Bkv = 512$ the measured average $vl$ is 512, the full architectural width, so vector length is not what limits this region. The access pattern is. Computing $\mathbf{Q}_i\mathbf{K}_j^{\!\top}$ needs $\mathbf{K}_j$ transposed, and $\mathbf{K}$ is stored with $d$ contiguous, so walking the contraction axis steps by the row pitch, 256 bytes at $d = 64$, a separate cache line per element. Read in place, without packing, 88.9\% of the region's vector memory instructions are strided. A strided access costs 128--512 cycles at full length where a unit-stride one costs 32--128 (Table~\ref{tab:latency}), so the region pays for the class of its instructions, not their count.

Packing solves the strided case. Copying the key block into a contiguous buffer before the product leaves the arithmetic and the vector length untouched and replaces almost every strided load with a unit-stride one. It makes \gemmqk{} $2.38\times$ faster. Packing is performed by the matrix-multiplication routine the kernel calls, in every arm, so every result below includes it; it is long-established~\cite{goto2008anatomy,vanzee2015blis} and not the subject of this paper.
\subsection{Contiguous operands, short vectors: the value product}
\label{sec:char:starve}

\gemmpv{} has no access-pattern problem. Its first operand was written by the kernel moments earlier and is unit-stride by construction; its second, $\mathbf{V}_j$, is read along the contiguous head dimension. Both operands are already contiguous along the axis the product sweeps, so packing has nothing to offer here. What limits the region is that the product is untransposed, which makes $d$ its vector axis: its instructions request 64 elements from registers that hold 512, so the region runs at an eighth of the architectural width whatever the block size, and no block size changes $d$. Against \gemmqk{} at the same tile this is the cleanest contrast available: two dense products over the same tile, differing in which axis each sweeps. The width is fixed by the loop shape, so the remedy is a different axis, which Section~\ref{sec:reorient} supplies.

\section{A Long-Vector Softmax}
\label{sec:softmax}
This section leaves the traversal of Section~\ref{sec:char} alone and works on the softmax. The region evaluates an exponential in software, and its cost on this machine turns out to be the polynomial's coefficients rather than its instructions. Two changes follow: the coefficients move into scalar registers, and the three passes over a score row fold into one. Section~\ref{sec:softmax:ceiling} gives what they are worth together and the limit they run into.
\subsection{How the exponential is computed}
\label{sec:softmax:exp}

The processor has no floating-point exponential, so the softmax evaluates one in software, per element. The standard construction splits the argument into a part that is exact in binary and a part small enough for a short polynomial~\cite{muller2016elementary}. Writing
\begin{equation}
  \label{eq:rangered}
  x = k\ln 2 + r, \qquad k = \left\lfloor x/\ln 2 \right\rceil \in \mathbb{Z},
  \qquad \lvert r \rvert \le \tfrac{\ln 2}{2},
\end{equation}
gives $e^{x} = 2^{k}e^{r}$. The first factor is free: $2^{k}$ is an FP32 number whose mantissa is zero, so it is built by adding the exponent bias to $k$ and shifting it into the exponent field, which is integer work. Over that interval a degree-6 polynomial approximates the second to single precision, and Horner's rule~\cite{knuth1997taocp2,higham2002accuracy} evaluates it as a chain of multiply--adds,
\begin{equation}
  \label{eq:horner}
  p \leftarrow p\,r + c_j, \qquad j = 5, 4, \ldots, 0 ,
\end{equation}
starting from $p = c_6$. Algorithm~\ref{alg:exp} gives the whole routine for one vector of $vl$ scores, its lines 1--3 the range reduction, 4--6 the Horner chain and 7--8 the $2^{k}$ rebuild; bold is a vector register throughout and italic a scalar. Counting the multiply--add that forms $r$, an exponential is seven multiply--adds, six of them a dependent chain. Every arm below uses this same degree-6 polynomial, so every comparison is at equal accuracy. On this machine the seven \emph{coefficients}, not the seven instructions, turn out to be the expensive part.

\begin{algorithm}[t]
\caption{The software exponential.}
\label{alg:exp}
\begin{algorithmic}[1]
\Require $\mathbf{x} \in \mathbb{R}^{vl}$, coefficients $c_0 \ldots c_6$
\State $\mathbf{x} \gets \max(\mathbf{x}, -87)$
       \Comment{keeps $2^{k}$ in exponent range}
\State $\mathbf{k} \gets \mathrm{round}(\mathbf{x}\log_2 e)$
       \Comment{to integer}
\State $\mathbf{r} \gets \mathbf{x} - \mathbf{k}\ln 2$
       \Comment{one multiply--add}
\State $\mathbf{p} \gets c_6$
\For{$j = 5$ \textbf{down to} $0$}
  \State $\mathbf{p} \gets \mathbf{p}\,\mathbf{r} + c_j$
        \Comment{Equation~\eqref{eq:horner}}
\EndFor
\State $\mathbf{e} \gets \mathrm{reinterpret}\bigl((\mathbf{k} + 127) \ll
       23\bigr)$ \Comment{$2^{k}$, integer}
\State \Return $\mathbf{p}\,\mathbf{e}$
\end{algorithmic}
\end{algorithm}

\subsection{The constants were vectors}
\label{sec:softmax:consts}

The softmax region reads each score row to find its maximum, again to exponentiate and again to sum, then writes it back. The trace nevertheless counted $62\,416$ vector memory instructions for the region and 126.1\,MB moved, against $13\,264$ and 25.4\,MB once the coefficients were moved into scalar registers and nothing else changed. The difference is 24 memory instructions on every one of the 2{,}048 calls, and it was spill traffic.

Line 6 of Algorithm~\ref{alg:exp} reads as a multiply--add, and on a three-operand vector ISA the obvious encoding is the fused form, which takes all three operands in vector registers:
\begin{align}
  \label{eq:fma}
  \text{fused:}\quad
    &\mathbf{p} \leftarrow \texttt{vfmadd.vv}(\mathbf{p},\, \mathbf{r},\,
      \mathbf{c}_j) \\
  \label{eq:split}
  \text{split:}\quad
    &\mathbf{p} \leftarrow \texttt{vfadd.vf}\bigl(
      \texttt{vfmul.vv}(\mathbf{p},\, \mathbf{r}),\, c_j \bigr)
\end{align}

Equation~\eqref{eq:fma} is one instruction against two, and with abundant vector registers it is the better encoding. Here it is not: the seven coefficients held that way are seven registers live across the whole polynomial, competing with the tile data and the running state for the 32 the compiler allocates, and the allocator spills~\cite{chaitin1982register}. The spilled constants are reloaded on every call, which is the excess the trace had found. No single instruction can take the coefficient from a scalar register instead, because the scalar operand of every \texttt{.vf} multiply--add (\texttt{vfmacc}, \texttt{vfmadd} and their negated forms) is a multiplicand, while Horner's rule needs $c_j$ as the addend. Equation~\eqref{eq:split} therefore pays one extra instruction per step to keep every coefficient in the scalar file, which is separately managed and not under pressure.

A shorter polynomial could relieve the pressure. Degree 4 holds five coefficients rather than seven and may fit, but it pays in accuracy and leaves the mechanism in place, since each remaining $c_j$ is still a live vector register. Equation~\eqref{eq:split} leaves the accuracy alone and holds none of them, at one extra instruction per step.

\begin{table}[t]
  \centering
  \caption{The softmax region across the measured changes, at
           $\Bq = 128$, $\Bkv = 512$, $N = 1024$, $d = 64$.}
  \label{tab:softmax}
  \footnotesize
  \setlength{\tabcolsep}{4pt}
  \begin{tabular}{@{}lrrr@{}}
    \toprule
    & Stock & Scalar & Fused \\
    & kernel & coeffs. & traversal \\
    \midrule
    Vector memory inst.     & 62{,}416  & 13{,}264  & 9{,}168 \\
    Vector arithmetic inst. & 45{,}544  & 59{,}880  & 41{,}448 \\
    Other vector inst.      & 41{,}448  & 37{,}352  & 16{,}872 \\
    Total vector inst.      & 149{,}408 & 110{,}496 & 67{,}488 \\
    Moved bytes (MB)        & 126.1     & 25.4      & 17.0 \\
    \midrule
    Region cycles (M)       & 19.92     & 12.96     & 8.00 \\
    \bottomrule
  \end{tabular}
\end{table}

Table~\ref{tab:softmax} records the outcome. From the stock kernel to scalar coefficients, vector memory instructions fall by 78.7\% and moved bytes by a factor of five, while vector arithmetic rises and the region runs $1.54\times$ faster in cycles.

\subsection{A fused traversal removes what remains}
\label{sec:softmax:fuse}

With the spills gone, the region's remaining traffic is its own structure: it reads each row of the score tile to find its maximum, again to exponentiate, and again to sum. Holding the row in a vector register across all three passes is possible only because a row of $\Bkv \le \vlmax$ elements fits in a single register, so the maximum, the exponential and the sum all see it without a reload. Memory instructions fall by a further 30.9\%, the two loads per call the algorithm predicts, and moved bytes by a third. Arithmetic falls by 9 instructions per call and the other classes by 10, almost all of them moves and configuration changes (Table~\ref{tab:softmax}). The region's cycles fall by $1.62\times$.

\subsection{The ceiling}
\label{sec:softmax:ceiling}

Together the two changes reduce the softmax region's cycles by $2.49\times$ and the whole kernel's by $1.27\times$. The softmax is then a fifth of the budget, so removing it entirely would buy at most $1.24\times$. Further work inside the softmax is not where the remaining time is.

The fused traversal also carries a precondition. Folding three passes into one keeps a row of the score tile in a register between them, so it requires $\Bkv \le \vlmax$ and the saving disappears above that. The natural layout's best key block is 512, exactly \vlmax{} on this machine, so the optimisation applies at the measured optimum and not beyond it. The transposed form of Section~\ref{sec:reorient} has no such precondition: its vector length is $\Bq$, it accumulates down the key rows, and no row of the tile is ever held resident.


\begin{figure*}[t]
  \centering
  \includegraphics[width=\textwidth]{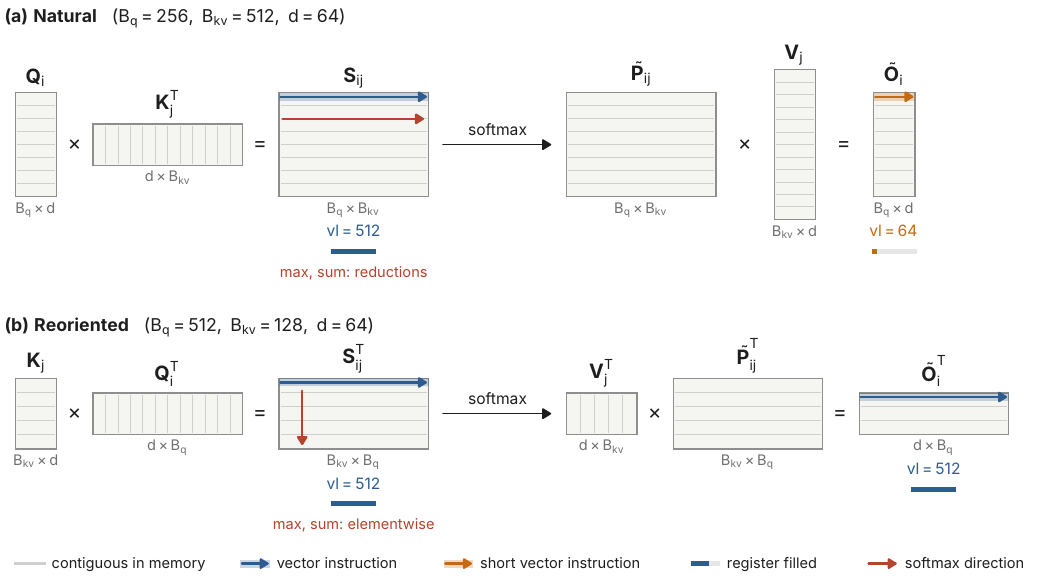}
  \caption{One tile of the forward pass in each orientation, each at its own
           best tile. Grey lines mark the contiguous axis in memory; blue
           arrows mark a full-length vector instruction, orange a short one,
           and red the direction the softmax combines. The bar under each
           output shows how much of the register its instructions fill. Sizes
           are to a compressed scale.}
  \label{fig:axes}
\end{figure*}

\section{Reorienting the Vector Axis}
\label{sec:reorient}

This section transposes every term of the recurrence. Section~\ref{sec:reorient:form} gives the transposed form and what changes in each region. Section~\ref{sec:reorient:softmax} treats the softmax, where the change is structural rather than a matter of width: two classes of instruction leave the region. Section~\ref{sec:reorient:model} gives an instruction-count model of when the transpose wins.

\subsection{The transposed formulation}
\label{sec:reorient:form}

At the natural layout's own best tile, $\Bkv = 512$, \gemmqk{} and the softmax already run at the register width (Table~\ref{tab:axis}). \gemmpv{} cannot: it takes its length from $d$ (Section~\ref{sec:char:starve}), and its average $vl$ stays near 64 at every tile in Table~\ref{tab:axis}.

What decides which axis each region sweeps is the order in which the kernel walks the tile, and Algorithm~\ref{alg:natural} never says what that order must be. So we transpose the two matrices the kernel builds:
\begin{equation}
  \label{eq:transposed}
  \mathbf{S}^{\!\top}_{ij} =
    \alpha\,\mathbf{K}_j\mathbf{Q}_i^{\!\top}, \qquad
  \tilde{\mathbf{O}}^{\!\top}_i =
    \mathbf{V}_j^{\!\top}\tilde{\mathbf{P}}^{\!\top}_{ij}.
\end{equation}
The algorithm is unchanged: every product accumulates in the same order, and the softmax computes the same maximum and sum, now vertically, down the key rows of the tile (Section~\ref{sec:reorient:softmax}). Figure~\ref{fig:axes} shows one tile in each orientation. The new shapes change four things.

The score tile is now $\Bkv \times \Bq$, so a row of it holds queries instead of keys. That is where the vector length now comes from.

$\mathbf{K}_j$ is now read in the orientation it is stored in, and $\mathbf{Q}_i$ is the block read across the stride. So $\mathbf{Q}_i$ is what the matrix-multiplication routine packs, on every call as it packed $\mathbf{K}_j$ before, and the strided loads of packing now scale with $\Bq$ instead of $\Bkv$.

We never build $\mathbf{V}_j^{\!\top}$. Transposing a product swaps which operand supplies the vector axis, and neither operand has to move for that, so the value product reads $\mathbf{V}_j$ where it already sits and sweeps $\Bq$.

The output comes out transposed, and putting it back is almost free. The kernel already walks the output once to divide by $\boldsymbol{\ell}_i$, so that pass now reads unit-stride and writes strided instead. It adds 9 to 13\% to the normalisation region, which is under half a percent of the kernel.

This change is outer-loop vectorisation~\cite{nuzman2008outer}: the query loop carries the softmax state, and making it the vector loop turns that state into a vector. With 512 elements in a register, carrying it as a vector is exactly what removes the reductions. Algorithms~\ref{alg:natural} and~\ref{alg:reor} are the same kernel in the two orientations. One parameter, $\Bq$, now sets the vector length of every region, replacing $\Bkv$ and $d$. The two \emph{reduction} lines of the first become the two \emph{elementwise} lines of the second. Table~\ref{tab:axis} shows this and measures it.

The bottom half of Table~\ref{tab:axis} matters as much as the top. At $\Bq = 128$ the reoriented kernel runs every region at 128, which is worse than the 512 the natural layout reaches in two of its three regions. Reorientation does not make the kernel faster at every tile. It puts the vector length of every region under one parameter, $\Bq$, and everything then depends on setting that parameter to the register width. At $\Bq = 128$ the reoriented kernel is slower than the fused one; at $\Bq = 512$, the register width, it is $1.43\times$ faster.

\begin{algorithm}[t]
\caption{Reoriented (transposed) FlashAttention-2 forward.}
\label{alg:reor}
\small
\begin{algorithmic}[1]
\Require $\mathbf{S}^{\!\top}_{ij}$ is $\Bkv \times \Bq$;\;
         $\mathbf{m}_i, \boldsymbol{\ell}_i$ are vectors over the query block
\For{each query block $i$}
  \State $\tilde{\mathbf{O}}^{\!\top}_i \gets 0$;\;
         $\mathbf{m}_i \gets -\infty$;\;
         $\boldsymbol{\ell}_i \gets 0$
  \For{each key block $j$}
    \State $\mathbf{S}^{\!\top}_{ij} \gets \mathbf{K}_j \mathbf{Q}_i^{\!\top}$
          \Comment{$vl = \Bq$}
    \State $\mathbf{m}' \gets
            \max\bigl(\mathbf{m}_i, \max_r \alpha\mathbf{S}^{\!\top}_{ij}[r,:]\bigr)$
          \Comment{\emph{elementwise}}
    \State $\corr_i \gets \exp(\mathbf{m}_i - \mathbf{m}')$;\;
           $\mathbf{m}_i \gets \mathbf{m}'$
          \Comment{one vector $\exp$}
    \State $\mathbf{S}^{\!\top}_{ij} \gets
            \exp(\alpha\mathbf{S}^{\!\top}_{ij} - \mathbf{m}_i)$
          \Comment{$vl = \Bq$}
    \State $\boldsymbol{\ell}_i \gets \corr_i \boldsymbol{\ell}_i +
            \textstyle\sum_r \mathbf{S}^{\!\top}_{ij}[r,:]$
          \Comment{\emph{elementwise}}
    \State $\tilde{\mathbf{O}}^{\!\top}_i \gets
            \corr_i \odot \tilde{\mathbf{O}}^{\!\top}_i
            + \mathbf{V}_j^{\!\top}\mathbf{S}^{\!\top}_{ij}$
          \Comment{$vl = \Bq$}
  \EndFor
  \State $\mathbf{O}_i \gets
          (\tilde{\mathbf{O}}^{\!\top}_i / \boldsymbol{\ell}_i)^{\!\top}$
        \Comment{transpose folded in}
\EndFor
\end{algorithmic}
\end{algorithm}

\begin{table}[t]
  \centering
  \caption{Vector axis of each region and its average $vl$, under emulation at
           $d = 64$. Averages are over every vector instruction in the region,
           so the few issued at another length move them slightly off the
           axis length.}
  \label{tab:axis}
  \footnotesize
  \begin{tabular}{@{}lcc cc@{}}
    \toprule
    & \multicolumn{2}{c}{Vector axis}
    & \multicolumn{2}{c}{Average $vl$} \\
    \cmidrule(lr){2-3}\cmidrule(lr){4-5}
    Region & Natural & Reor. & Natural & Reor. \\
    \midrule
    \multicolumn{5}{@{}l}{\itshape $\Bq = 512$, $\Bkv = 128$} \\
    \gemmqk{}  & $\Bkv$ & $\Bq$ & 132.5 & \textbf{512.0} \\
    \softmaxr{}& $\Bkv$ & $\Bq$ & 132.0 & \textbf{512.0} \\
    \gemmpv{}  & $d$    & $\Bq$ & 66.7  & \textbf{512.0} \\
    \midrule
    \multicolumn{5}{@{}l}{\itshape $\Bq = 128$, $\Bkv = 512$} \\
    \gemmqk{}  & $\Bkv$ & $\Bq$ & \textbf{512.0} & 132.5 \\
    \softmaxr{}& $\Bkv$ & $\Bq$ & \textbf{505.5} & 128.1 \\
    \gemmpv{}  & $d$    & $\Bq$ & 64.7           & 128.6 \\
    \bottomrule
  \end{tabular}
\end{table}

\subsection{Softmax becomes vertical}
\label{sec:reorient:softmax}

The softmax is where the change is structural rather than merely wider: two classes of instruction leave the region altogether. In the natural orientation a query row occupies $\Bkv$ contiguous elements, so $\mathrm{rowmax}$ and $\mathrm{rowsum}$ are reductions \emph{along} a vector register and the running state $\mathbf{m}_i, \boldsymbol{\ell}_i$ is a pair of scalars per row. Transposed, a key row occupies $\Bq$ contiguous elements and the same two quantities are indexed by the query axis, so they become vector registers and the reductions become elementwise accumulations down the key rows. The reoriented softmax makes two passes over the tile,
\begin{equation}
  \label{eq:vertical}
  \mathbf{m}' \leftarrow \max\!\left(\mathbf{m}',\,
      \alpha\mathbf{S}^{\!\top}_{ij}[r,:]\right), \quad
  \mathbf{s} \leftarrow \mathbf{s} +
      \exp\!\left(\alpha\mathbf{S}^{\!\top}_{ij}[r,:] - \mathbf{m}'\right),
\end{equation}
the first for $r = 0, \ldots, \Bkv - 1$ starting from $\mathbf{m}' = \mathbf{m}_i$, the second over the same rows starting from $\mathbf{s} = 0$ and storing each exponentiated row, after which $\boldsymbol{\ell}_i \leftarrow \corr_i\boldsymbol{\ell}_i + \mathbf{s}$. Every operation is elementwise over the $\Bq$ elements of a row, one \texttt{vsetvli} configures the whole tile, and $\mathbf{m}_i$ and $\boldsymbol{\ell}_i$ stay in vector registers across all $\Bkv$ rows. Figure~\ref{fig:axes} shows the two directions.

Two costs disappear and a third shrinks. The two reductions per query row per key block are gone and so is the move that pulled each reduction's result into a scalar register. The third is $\corr_i$. Both forms compute it once per query row per key block, but the natural orientation holds the running maximum as one scalar per row and evaluates each $\corr_i$ as a separate scalar exponential, while the transposed form holds it as a vector and covers a whole query block with one vector exponential.

One cost is added. The two passes read each score twice where the fused traversal read it once, which at each arm's own best tile is 10\% more memory instructions in the region.

\subsection{When the transpose wins}
\label{sec:reorient:model}

The transpose changes no arithmetic. It changes how many vector instructions carry it. For a block no longer than \vlmax{} a row is one vector, so a region's instruction count is the number of rows it walks times what it issues per row. Each region then compares in its own way, natural over reoriented.

\emph{Value product.} Both forms issue the same instructions per row, but the natural form's rows are $d$ long and the reoriented form's $\Bq$, so the ratio is
\begin{equation}
  \label{eq:ratios}
  \frac{\Bq}{d}.
\end{equation}
This is the only place the head dimension enters. The transpose wins here whenever $\Bq > d$, and the gain falls as $d$ grows.

\emph{Score product.} Again the instructions per row are the same; the rows are $\Bkv$ long in one form and $\Bq$ in the other, so the ratio is $\Bq/\Bkv$. It is a trade of one block length for the other and is even when the two are equal.

\emph{Softmax.} The natural form walks $N^2/\Bkv$ query rows of $\Bkv$ elements, the reoriented form $N^2/\Bq$ key rows of $\Bq$. Per row both do the same elementwise work $E$: scale, maximum, subtract, exponential, accumulate and store. Both rescale the accumulator, which is the value product ratio again and is counted there. Both compute $\corr_i$ once per row, in scalar code in one form and as one vector exponential per tile in the other, and neither is counted here.

What differs is the price of the maximum and the sum. In the natural orientation a row is one vector, so both are reductions, one \texttt{vfredmax} and one \texttt{vfredusum}, each followed by a move to a scalar register. Transposed, a row holds different queries, so the maximum is a \texttt{vfmax} and the sum a \texttt{vfadd} down the key rows, and nothing leaves the vector registers. The transpose pays two extra memory instructions per row instead, because it reads the score tile a second time and writes the scaled scores back before it:
\begin{equation}
  \label{eq:softmaxratio}
  \underbrace{E + 2R + 2M}_{\text{natural}}
  \qquad\text{against}\qquad
  \underbrace{E + 2U}_{\text{reoriented}},
\end{equation}
with $R$ a reduction, $M$ a vector to scalar move and $U$ a unit-stride access.

The transpose wins the row whenever $2U < 2R + 2M$. Table~\ref{tab:latency} prices a reduction at 160 cycles and a unit-stride access at 32 to 128, so the row gives up at most 256 cycles and saves at least 320. That margin assumes the second read hits in cache. The tile was written by the pass immediately before, but nothing guarantees it survives. Over the whole kernel the natural form issues $2N^2/\Bkv$ reductions and the reoriented form none. No timing in this paper is derived from the model.
\section{Methodology}
\label{sec:method}

\subsection{Target platform}
\label{sec:method:target}

EPAC~\cite{mantovani2026epac} was developed within the European Processor Initiative: an in-order RISC-V scalar core driving a decoupled vector unit whose registers hold $16\,384$ bits. Table~\ref{tab:platform} lists the parameters the kernel design depends on, and four of them shape everything below.

\emph{Addressing is at cache-line granularity}. A unit-stride access uses every element of each 64-byte line it brings in. An access whose stride is larger than a line uses one element per line, so it brings in sixteen times more data than it consumes. \emph{There is no operand chaining}, so a dependent instruction waits until its producer has retired every element and the cost of a sequence tracks its dependence depth, at the latencies of Table~\ref{tab:latency}. \emph{The register file is small in registers and large in bytes}: the compiler allocates 32 architectural registers and 40 physical registers, so a kernel holding too many live values spills~\cite{chaitin1982register}, and one spilled register at full length is 2\,KB of traffic. \emph{There is no hardware exponential}, so the softmax evaluates a software polynomial whose cost is its instruction count and dependence structure, both under the kernel's control.

The vector unit implements RVV 0.7.1~\cite{riscvv2019v07}, the draft current when it was designed; a successor implements the ratified 1.0~\cite{mantovani2026epac}. Nothing below depends on the difference: every instruction the kernels use has a 1.0 counterpart, and the limits Section~\ref{sec:char} identifies come from loop lengths, which no 1.0 feature changes.

\begin{table}[t]
  \centering
  \caption{EPAC parameters the kernel design depends on.}
  \label{tab:platform}
  \footnotesize
  \begin{tabular}{@{}ll@{}}
    \toprule
    Parameter & Value \\
    \midrule
    ISA, vector extension      & RV64GCV, RVV 0.7.1 \\
    Scalar core                & In-order, dual-issue, 4 ports \\
    Vector unit                & Decoupled, no operand chaining \\
    Lanes                      & 8, one FMA each, 2 FP64 FLOP/cycle \\
    Peak throughput            & 32 FP32 FLOP/cycle \\
    Element widths             & 64 and 32\,b; no 16\,b \\
    Maximum vector length      & $16\,384$\,b, 512 FP32 \\
    Vector registers           & 32 architectural, 40 physical \\
    Lane to memory             & 64\,b/cycle each \\
    L1 D-cache                 & 32\,KB, 64\,B line \\
    L2 cache                   & 1\,MB, 4 banks, 8-way, inclusive \\
    Hardware FP exponential    & None \\
    \midrule
    Clock                      & chip 1\,GHz; FPGA 50\,MHz \\
    Sustained memory bandwidth & chip 6--7\,B/cycle; FPGA 32\,B/cycle \\
    Interconnect ceiling       & 64\,B/cycle \\
    \bottomrule
  \end{tabular}
\end{table}

\begin{table}[b]
  \centering
  \caption{Vector instruction latency, as documented by the vendor at
           $vl = 512$ FP32 elements and at no other length.}
  \label{tab:latency}
  \footnotesize
  \begin{tabular}{@{}llr@{}}
    \toprule
    Class & Instructions & Cycles \\
    \midrule
    Strided memory      & \texttt{vlse}, \texttt{vsse}           & 128--512 \\
    FP reduction        & \texttt{vfredsum}, \texttt{vfredmax}   & 160 \\
    Unit-stride memory  & \texttt{vle}, \texttt{vse}             & 32--128 \\
    Moves               & \texttt{vmv.v.x}, \texttt{vmv.v.v}     & 50 \\
    Arithmetic          & \texttt{vfadd}, \texttt{vfmadd}        & 35 \\
    \bottomrule
  \end{tabular}
\end{table}

\subsection{Measurement targets and instrumentation}
\label{sec:method:platforms}

Cycle counts come from two targets of the same design, the \chip{} at 1\,GHz and an FPGA implementation of the same design at 50\,MHz (Table~\ref{tab:platform}). Both implement the same processor revision, so a change that helps on one is expected to help on the other; they differ in clock and in sustained memory bandwidth per cycle. The 64\,B/cycle ceiling is fixed by the cache-line interconnect; the \fpga{} sustains about half of it, while on the chip the SerDes link to the DDR4 bank limits at 1\,GHz and sustains roughly a tenth. The chip therefore charges far more for memory traffic relative to arithmetic than the \fpga{} does, which Section~\ref{sec:eval:platforms} measures directly. No result below is averaged across the targets and no table mixes them; results lead with the chip. Every arm is vectorised, so no figure below is a speed-up over scalar code. The absolute reference is the machine's peak of 32 FP32 FLOP/cycle, against which the stock kernel reaches 16.2\%. All measurements are FP32, on one core and one attention head, at $d = 64$ and $N = 1024$ unless stated.

Instruction-level data comes from the RAVE QEMU plugin~\cite{vizcaino2025rave}, which reports per-region instruction mixes, effective vector lengths and moved bytes. It models register width only: no caches, no pipeline, no latencies. Its role is to establish which instruction stream the compiler and the intrinsics emit, which the hardware counters cannot show. Cycles and retired instructions are read per region through a tracing interface~\cite{pillet1995paraver}, each reported against total cycles for the same region, because an absolute instruction count can move opposite to performance when a kernel does less work in a denser issue stream.

\subsection{Region attribution}
\label{sec:method:regions}

Counters and cycles are attributed to named regions rather than to the kernel as a whole, because the regions have different bottlenecks and a whole-kernel number cannot separate them.Four are instrumented:

\begin{itemize}
  \item \gemmqk{}, the score product $\mathbf{Q}_i\mathbf{K}_j^{\!\top}$;
  \item \softmaxr{}, which applies $\alpha$, updates $\mathbf{m}_i$,
        exponentiates, sums the rows and forms $\corr_i$;
  \item \gemmpv{}, the value product $\tilde{\mathbf{P}}_{ij}\mathbf{V}_j$;
  \item \texttt{output\_scaling}, the normalisation
        $\tilde{\mathbf{O}}_i / \boldsymbol{\ell}_i$.
\end{itemize}

\noindent The first three run once per $(i,j)$ tile and the last once per query block, so each carries an execution count and per-tile costs follow by division. The scale $\alpha$ is applied inside \softmaxr{}, so \gemmqk{} is an unscaled product. Cycles outside the four appear as \emph{other} and are under 1\% in every arm. Boundaries sit inside the kernel and are identical in every arm, independent of which matrix-multiplication routine a region calls. All measurements are FP32, the narrowest type the vector unit computes in, on one core, at $d = 64$ and $N = 1024$ unless stated.

\section{Evaluation}
\label{sec:eval}

\subsection{End-to-end performance}
\label{sec:eval:e2e}

\begin{figure}[t]
  \centering
  \includegraphics[width=\columnwidth]{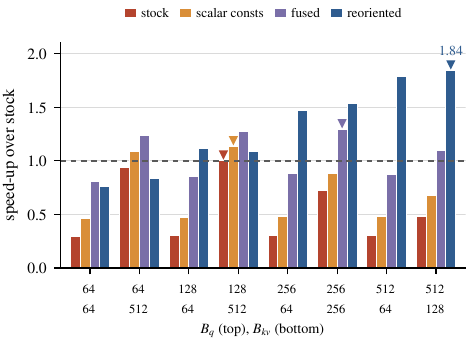}
  \caption{Speed-up over the stock kernel at its own fastest tile, \chip{},
           $N = 1024$, $d = 64$. Shades darken from the stock to the reoriented
           kernel; the fused and reoriented arms each build on scalar
           coefficients. A triangle marks each arm's own fastest tile.}
  \label{fig:tiles}
\end{figure}

Figure~\ref{fig:tiles} gives every arm at each of the eight tile shapes measured, as a speed-up over the stock kernel at its own fastest tile, so a bar reads as what replacing a tuned stock kernel buys. The reoriented kernel is fastest at $\Bq = 512$, $\Bkv = 128$, and the stock and scalar-coefficient kernels at $\Bq = 128$, $\Bkv = 512$. The fused kernel is nearly flat across the two, 40.23\,Mcyc at $\Bq = 256$, $\Bkv = 256$ against 40.88 at $\Bq = 128$, $\Bkv = 512$. The arms branch rather than stack: the fused and reoriented kernels both build on scalar coefficients, and reorientation replaces the fused traversal instead of adding to it. Quoting each arm where it is fastest (Table~\ref{tab:platforms}), scalar coefficients are worth $1.13\times$, adding the fused traversal $1.29\times$, and adding reorientation instead $1.84\times$. Against the fused kernel the reoriented one is $1.43\times$ faster, 40.23 against 28.22\,Mcyc. Floating-point utilisation rises from 16.2\% to 29.7\% of peak.

\begin{table}[b]
  \centering
  \caption{Each arm at its own fastest tile on each target, $N = 1024$,
           $d = 64$. The fused and reoriented arms each build on scalar
           coefficients. The two targets run at different clocks, so the factors
           compare and the cycles do not.}
  \label{tab:platforms}
  \footnotesize
  \begin{tabular}{@{}lrrrr@{}}
    \toprule
    & \multicolumn{2}{c}{\chip{}} & \multicolumn{2}{c}{\fpga{}} \\
    \cmidrule(lr){2-3}\cmidrule(lr){4-5}
    Arm           & Mcyc  & factor        & Mcyc  & factor \\
    \midrule
    stock           & 51.94 & $1.00\times$   & 39.37 & $1.00\times$ \\
    scalar consts   & 45.83 & $1.13\times$   & 30.18 & $1.30\times$ \\
    fused           & 40.23 & $1.29\times$   & 25.38 & $1.55\times$ \\
    reoriented      & 28.22 & $1.84\times$   & 14.55 & $2.71\times$ \\
    \midrule
    \% of peak    & \multicolumn{2}{c}{16.2 to 29.7}
                  & \multicolumn{2}{c}{21.3 to 57.7} \\
    \bottomrule
  \end{tabular}
\end{table}

\subsection{The two targets rank the changes identically}
\label{sec:eval:platforms}

Table~\ref{tab:platforms} puts the \fpga{} beside the chip, each arm at its own fastest tile on each. Every step is larger on the \fpga{}, uniformly and by a growing margin, while the ordering and the winner are the same on both.

The cause is in the measurement rather than inferred. Table~\ref{tab:ratio} divides chip cycles by \fpga{} cycles for the same arm, tile and region, which says how much more each kind of work costs on silicon. Two regions bracket the rest. \gemmqk{} is almost all arithmetic and costs about $1.6\times$ more on the chip, for all four arms. The normalisation pass is almost all memory traffic and costs about $3.7\times$ more, likewise for all four.

The softmax walks across that bracket as it is optimised, from $1.17\times$ for the stock kernel to $2.98\times$ reoriented. Each change removes vector instructions and spill traffic that both machines charge for, and what is left behind is a read and a write of the score tile, which the chip charges three times as much for. The better the kernel, the more of its remaining cost is memory and the more the two targets diverge. That is why the \fpga{} scores every step higher and why the chip figure is the one to quote.

\begin{table}[b]
  \centering
  \caption{Chip cycles per \fpga{} cycle, same arm, same region, both at
           $\Bq = 256$, $\Bkv = 512$. Arithmetic and memory traffic bracket
           the rest.}
  \label{tab:ratio}
  \footnotesize
  \begin{tabular}{@{}lrrrr@{}}
    \toprule
    Region & stock & scalar & fused & reor. \\
    \midrule
    \gemmqk{}               & 1.62 & 1.60 & 1.58 & 1.56 \\
    \softmaxr{}             & 1.17 & 1.57 & 2.33 & 2.98 \\
    \gemmpv{}               & 1.46 & 1.45 & 1.42 & 2.64 \\
    \texttt{output\_scaling}& 3.75 & 3.65 & 3.65 & 3.84 \\
    \bottomrule
  \end{tabular}
\end{table}

\subsection{Where the gain comes from}
\label{sec:eval:gain}

\begin{figure}[t]
  \centering
  \includegraphics[width=\columnwidth]{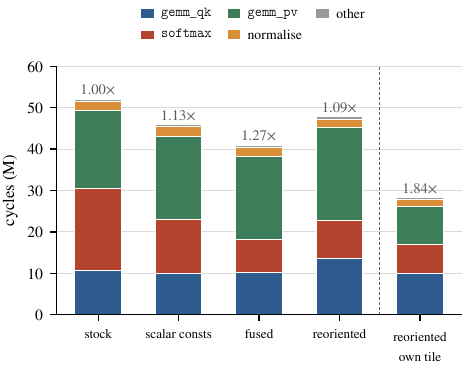}
  \caption{Cycles by region on the \chip{}, $N = 1024$, $d = 64$. The first
           four bars share $\Bq = 128$, $\Bkv = 512$, the fastest tile of all
           three natural-layout arms; the fifth is the reoriented kernel at
           its own fastest tile. Each region is limited by a different bottleneck: the value product by vector length, the softmax by its
           instruction classes, the score product by neither.}
  \label{fig:regions}
\end{figure}

Figure~\ref{fig:regions} breaks the arms down by region. The first four bars share $\Bq = 128$, $\Bkv = 512$, the fastest tile of all three natural-layout arms, so for those three the comparison is at a common block size and at each one's own optimum at once. At that tile the reoriented kernel runs at a quarter of the register width and is not competitive; the fifth bar is the same kernel at its own tile. The gain has a different source in each region.

\emph{Value product: vector length.} \gemmpv{} takes its length from $d$ in the natural form and from $\Bq$ in the reoriented one, so the ratio is $\Bq/d$ (Equation~\eqref{eq:ratios}). Against the fused kernel, each at its own best tile, the region is $2.01\times$ faster reoriented. This is the only region whose ratio contains the head dimension, and the one that carries most of the whole-kernel gain.

\emph{Softmax.} At each arm's own best tile the region is $1.37\times$ faster reoriented. The reoriented form issues no reductions and no vector-to-scalar moves, and it computes $\corr_i$ as one vector exponential per tile instead of one scalar exponential per query row.

\emph{Score product: neither.} \gemmqk{} issues the same instructions per row in both orientations, so it follows $\Bq/\Bkv$ and is within 1\% of even at the optima, where both block lengths are 512.

\emph{Sensitivity to tiling.} The worst-to-best ratio over the eight measured tiles is $3.42\times$ for the stock kernel and $2.42\times$ for the reoriented one. Reorientation makes the kernel about $1.4\times$ less sensitive to the tile, but that was not its purpose. What it changes is which parameter the kernel is sensitive to: the natural layout takes its vector length from $\Bkv$ and is slowest where the key block is short, while the reoriented one takes it from $\Bq$ and is fastest there. The inversion is visible straight across Figure~\ref{fig:tiles}.

\subsection{How the gain varies}
\label{sec:eval:scaling}

\begin{table}[t]
  \centering
  \caption{How the gain varies, \chip{}, each arm at its own best tile.}
  \label{tab:vary}
  \footnotesize
  \begin{tabular}{@{}lrrr@{}}
    \toprule
    & \multicolumn{2}{c}{cycles per score} & \\
    \cmidrule(lr){2-3}
    Setting & stock & reor. & speed-up \\
    \midrule
    \multicolumn{4}{@{}l}{\itshape Sequence length, $d = 64$} \\
    $N = 1024$ & 49.5 & 26.9 & $1.84\times$ \\
    $N = 2048$ & 47.7 & 23.6 & $2.02\times$ \\
    $N = 4096$ & 46.8 & 22.5 & $2.08\times$ \\
    \bottomrule
  \end{tabular}
\end{table}

Table~\ref{tab:vary} varies, one at a time, the quantities the gain depends on.

\emph{Sequence length.} Attention is the part of a layer whose cost is quadratic in $N$, so what decides whether the result matters is whether it survives at long context. The speed-up does not decay: it is $1.84$ to $2.08\times$ over the measured range, and the two kernels keep the same optimal tiles, as Equation~\eqref{eq:ratios} predicts, since neither ratio contains $N$.

\emph{Head dimension.} The head dimension enters one region, \gemmpv{}, through the ratio $\Bq/d$ of Equation~\eqref{eq:ratios}. At $d = 128$ the reoriented kernel's advantage in that region falls from $2.01\times$ to $1.26\times$.

\section{Discussion}
\label{sec:disc}

\emph{What the result is conditional on.} Equations~\eqref{eq:ratios} and~\eqref{eq:softmaxratio} state the condition: the reoriented form wins in the value product whenever the query block is longer than the head dimension, in the score product once it is also longer than the key block. The width is what makes the orientation a first-order choice. On a short register every block spans many vector instructions, so the length ratios tend to one and the reductions are a small share of the instruction stream. The result therefore scales with register width: it is largest on the longest machines, and those are the machines on which attention is now expected to run. The result also assumes a long query block. It concerns prefill and training, where $\Bq$ can be set to \vlmax{}, and not autoregressive decode, where a query block holds one row or one query group and the natural orientation is the right one.

\emph{Where the remaining headroom is.} The reoriented kernel reaches 29.7\% of peak, so a factor of three is unclaimed, and the regions do not share it evenly. \gemmpv{} is at 46.4\% of peak and \gemmqk{} is now the largest single region at 35\% of the budget, while the softmax is a quarter of it and removing it entirely would buy at most $1.33\times$. The score product is the next target. It is a dense product whose operands are already packed, so what it leaves on the table is in the matrix-multiplication routine itself: register blocking and an inner kernel tuned for a decoupled unit without chaining. That work applies to both orientations alike and so cannot flatter either.

\section{Related Work}
\label{sec:related}

\emph{Attention kernels.} FlashAttention~\cite{dao2022flashattention} removes the quadratic memory cost by tiling both operands and carrying the normalisation as an online recurrence~\cite{milakov2018online}; later versions improve parallelism on GPUs~\cite{dao2023flashattention2} and exploit asynchrony and low precision~\cite{shah2024flashattention3}. FlashAttention-4~\cite{zadouri2026flashattention4} meets the problem of Section~\ref{sec:softmax} on GPUs whose matrix units have outgrown their exponential units, computing part of the exponentials as a polynomial on the FMA pipes and skipping most rescalings of the running maximum. That line of work treats the traversal as fixed. CPU implementations target short-vector SIMD~\cite{shen2023efficient,fu2024meatten} and vectorise along the head dimension, where any axis fills a register.

\emph{Attention on RISC-V vector processors.} Titopoulos et al.~\cite{titopoulos2026vectorized} vectorise FlashAttention for RVV along the key axis, with the maximum and sum as vector reductions and a bit-level approximation of the exponential, and evaluate registers of 32 to 128 elements in simulation. FlashAttention-V~\cite{gupta2026flashattentionv} meets the mismatch this paper addresses from the other side: once the register outgrows the head dimension it packs several heads into one register, keeping the reductions, and so extends to decode, where a query block is a single row, at the cost of needing $\vlmax/d$ heads; it is evaluated in simulation up to 8192 bits and on a 256-bit board. Reorienting along queries needs no extra heads and removes the reductions but needs a long query block, so the two apply to different cases. VEXP~\cite{wang2025vexp} adds an exponential instruction to a RISC-V core rather than computing it in software.

\emph{Long-vector architectures.} RISC-V vector implementations span a wide range of register widths~\cite{lee2015hwacha,cavalcante2020ara,perotti2022newara,cavalcante2022spatz,minervini2023vitruvius,sifive2022x280}. SX-Aurora~\cite{komatsu2018sxaurora} and the processor studied here~\cite{mantovani2026epac} sit at the long end of that range.

\emph{Restructuring kernels for long vectors.} Sparse matrix products on long-vector machines needed a different blocking and accumulation structure rather than a better-tuned loop~\cite{lefevre2023spgemm}; there the vector length was too short because the data was sparse, here because the implementation chose the wrong axis of a tile that is large in both.

\section{Conclusion}
\label{sec:concl}

The FlashAttention recurrence fixes which values a kernel computes. It does not fix the orientation in which they are computed, and on a processor whose registers hold 512 single-precision elements that freedom is worth $1.43\times$ over a fused natural-orientation kernel sharing the same exponential and the same matrix-multiplication routine. Transposing the score tile and the running output puts every region under one parameter, the number of queries in a tile, and turns the softmax's per-row reductions into elementwise accumulations. With the exponential's coefficients moved out of the vector registers where they spilled, the kernel is $1.84\times$ - $2.08\times$ faster on EPAC silicon; an FPGA of the same design ranks every change identically.

\bibliographystyle{IEEEtran}
\bibliography{refs}

\end{document}